# Testing approximate normality of an estimator using the estimated MSE and bias with an application to the shape parameter of the generalized Pareto distribution

J. Martin van Zyl

**Abstract** In this work the normality in small samples of three often used estimators of the shape parameter of the generalized Pareto distribution are investigated. Normality implies that hypotheses can be tested with confidence in smaller samples. Often it is not easy to choose between estimators, based on the estimated MSE and bias using simulation studies and normality give the added advantage that hypotheses concerning the parameter can be tested in small samples. A confidence interval for the index of the S&P500 index is found by applying the results to estimators of the generalized Pareto distribution.

## 1 Introduction

Three well-known estimators of the shape parameter of a generalized Pareto distribution (GPD) are investigated to see if they can be used to test hypotheses assuming normality in small samples and also to get an indication at which samples sizes it is acceptable to assume normality. Often it can be proven that an estimator is asymptotically normally distributed but that is very vague and for some estimators this can mean large sample sizes of thousands of observations which are often not available in practical problems. The shape parameter and the inverse of it, the tail index, is important in risk analysis and used to check if data is heavy-tailed. The probability weighted moment (PWM) estimator, maximum likelihood (ML) estimator and the empirical Bayes estimator proposed by Zhang and Stephens (2009) will be investigated. Samples from the three-parameter will be simulated and transformed to the two-parameter GPD by subtracting the scale parameter estimated as the smallest observation in a sample. It was found that the distribution of the Zhang and Stephens estimator (2009) is closest to normally distributed in small samples and in smaller samples it can be used with confidence to test hypotheses. The ML estimator is only normally distributed in larger samples of say n=250 and above and the PWM estimator is not even close to normally distributed in samples of size n=500.

The Jarque–Bera test (Jarque and Bera, 1987) and the Lilliefors test for normality (Lilliefors, 1967) will both be applied. It should be noted that normality as such does not imply good results with respect to bias and mean squared error (MSE), but all three these estimators were shown to perform well (Zhang and Stephens, 2009), (de Zea Bermudez, Kotz, 2010). Some of the estimators are asymptotically normally distributed and the asymptotic variances known, but not for small samples (Hosking and Wallis, 1987). Using these results it is shown in an application in section 5 that with 95% certainty, that the index of the S&P500 is at least 2.7196, thus with a finite variance.

A new test which can be applied specifically in simulation studies where the true parameter is known is also proposed and used. This is an exact method although very simple, and under the hypothesis of normality will involve no approximations when performing tests.

When testing a hypothesis, under the true null-hypothesis where the true parameter is known, the hypothesis can be rejected if the underlying distribution of the data is not distributed according to the distribution of the sample on which the test is based. In simulation studies where the true parameters are known, this aspect can be used to test normality. A t-test is based on the assumption of normally distributed observations when calculating the test statistic. Consider a general estimation problem of a parameter $\theta$ in a simulation study, which is a parameter of a specified distribution.

After m samples of sizes n each simulated were simulated with true parameter $\theta_0$, m estimators: $\hat{\theta}_1,...,\hat{\theta}_m$ are available with mean $\bar{\theta}$ and estimated variance $S_{\hat{\theta}}^2 = (1/m)\sum_{j=1}^{m}(\hat{\theta}_j - \bar{\theta})^2$ and the estimated variance of $\bar{\theta}$ is $S_{\hat{\theta}}^2/(m-1)$. To test the normality of the estimator, the following statistic can be used:

$$t_{m-1} = (m-1)(\bar{\theta} - \theta_0)/\sqrt{S_{\hat{\theta}}^2}. \qquad (1)$$

If the estimators, $\hat{\theta}_1,...,\hat{\theta}_m$, are normally distributed, the t-statistic would be central t distributed with $v = m-1$ degrees of freedom. If this statistic is used to test the hypothesis $H_0: \theta = \theta_0$, it will be rejected with high probably only if the estimator is not normally distributed, because the hypothesis is true and the estimated variance correct. This is exact under the hypothesis of normality and the size of this statistic can also give an indication of the distance of the distribution of the estimator and the normal distribution. It will be shown in section 2, that this statistic can be written in terms of the estimated bias and MSE at sample size n, which means that previously published results where the MSE and bias were reported can be used to check normality of an estimator.

An approximate variance can be investigated if it can be used at a sample size where it was shown that the estimator is normally distributed. Say it is established or known that at sample size n an estimator is normally distributed, and there is an expression for the variance (maybe an asymptotic variance) say $\hat{\sigma}_a^2$. If the expression for the variance is correct, and m t-statistics are formed under the true null hypothesis $t_j = (\hat{\theta}_j - \theta_0)/\hat{\sigma}_{aj}, j = 1,...,m$, the m statistics should be approximately for n large be distributed normal with mean 0 and variance 1. If the same principle is applied the statistic

$$t^*_{m-1} = (m-1)\bar{t}/\sqrt{S_t^2}, \text{ with } \bar{t} = (1/m)\sum_{j=1}^{m} t_j, \quad S_t^2 = (1/m)\sum_{j=1}^{m}(t_j - \bar{t})^2,$$

should have a t distribution with m-1 degrees of freedom.

Since m would almost always be large in a simulation study, the statistic would be approximately normally distributed and can also be compared against a standard normal critical value. If the test statistic is not standard normal distributed, using the approximate variances, a bootstrap estimate of the variance of the estimator must be calculated to perform hypothesis testing in practical problems. Both of these aspects are important when testing hypotheses, the distribution and the variance of the statistic.

An application of using the t-statistic to test normality using previously published results concerning estimator of the GPD parameters, where the MSE and bias were reported, will also be given.

## 2 Testing normality of estimators using the MSE and bias

It is first shown that the t-statistic can be written in terms of the estimated mean square error (MSE) and bias calculated from the simulated estimation results. This form can be used to check normality of an estimator using the MSE and Bias. Consider again the m estimators as described in section 1. The performance of an estimator was tested on the basis of m simulated samples of size n each, resulting in m estimators $\hat{\theta}_1, ..., \hat{\theta}_m$ of the true parameter $\theta_0$. It will be assumed that the estimators are asymptotically unbiased. Let $\bar{\theta} = \sum_{j=1}^{m} \hat{\theta}_j / m$ denote the sample mean of $\hat{\theta}_1, ..., \hat{\theta}_m$, and the estimated variance of $\hat{\theta}$ by $S^2 = \hat{\sigma}^2(\hat{\theta}) = \sum_{j=1}^{m}(\hat{\theta}_j - \bar{\theta})^2/(m-1)$, and the variance of $\bar{\theta}$ is $\hat{\sigma}^2(\bar{\theta}) = S^2/m$.

The estimated bias and MSE using the true parameter are respectively $B(\hat{\theta}) = \bar{\theta} - \theta_0$, and $MSE = (1/m)\sum_{j=1}^{m}(\hat{\theta}_j - \theta_0)^2$. It follows that

$$\begin{aligned} mMSE &= \sum_{j=1}^{m}(\hat{\theta}_j - \theta_0)^2 \\ &= \nu S^2 + m(\theta_0 - \bar{\theta})^2, \quad \nu = m-1 \\ &= \nu S^2 + mB^2(\hat{\theta}). \end{aligned}$$

Thus $S^2 = \frac{m}{\nu}(MSE - (\theta_0 - \bar{\theta})^2)$ and $\hat{\sigma}^2(\bar{\theta}) = \hat{\sigma}^2(\sum_{j=1}^{n} \hat{\theta}_j / m) = S^2 / m$. If $\hat{\theta}_1,...,\hat{\theta}_m$ are normally distributed under the true hypothesis, then the statistic in (1) is exactly t distributed and can be written as

$$t = \frac{\bar{\theta} - \theta_0}{S(\bar{\theta})} = \frac{B(\hat{\theta})}{\sqrt{\{\frac{m}{\nu}(MSE - B^2(\hat{\theta}))\}/m}}$$
$$= \sqrt{\nu}B(\hat{\theta}) / \sqrt{(MSE - B^2(\hat{\theta}))}, \nu = m-1 \qquad (2)$$
$$= \sqrt{\nu}(\bar{\theta} - \theta_0)/S \sim t_\nu$$

The number of simulated samples m will mostly be large, more than say 500 samples generated, thus the t-statistic is approximately standard normally distributed with the test statistic $z \sim N(0,1)$, $m \approx \nu$, if the estimator is normally distributed and the test statistic is

$$z = \sqrt{\nu}(\bar{\theta} - \theta_0) / \hat{\sigma}(\hat{\theta}). \qquad (3)$$

Rejection of the hypothesis $H_0: \theta = \theta_0$, will imply with high probability that the sample is not normal distributed.

The approximate distribution of the statistic z where the $\hat{\theta}'s$ are not normally distributed and the variance of the true parameters known, attracted much attention and the most well known approximation in such a case is the Edgeworth expansion for large samples with

$$f(z) = \phi(z)\left(1 + \rho_3 H_3(z)/(6\sqrt{n}) + (3\rho_4 H_4(z) + \rho_3^2 H_6(z))/(72n)\right) + O(n^{-3/2}),$$

where n the sample size, $\rho_r = \kappa_r / \kappa_2^{r/2}$ are the standardized cumulants and $\kappa_r$ denotes the r-th cumulant. $H_r(z) = (-1)^r \phi^{(r)}(z)/\phi(z)$ denotes a Hermite polynomial of order r. Using these results a sample approximation of the distribution is (Kendall, Stuart and Ord, 1987):

$$f(z) = \phi(z)\left(1 + (\hat{\rho}_3 3 - 5\hat{\rho}_3^2)/(24n)\right) + 0(n^{-1}).$$

Expression of the series in terms of the t-distribution function was considered in the paper of Finner and Dickhaus (2010). This can be considered as a way to calculate the power for a specific estimator for large sample sizes. It can be seen that for large sample sizes the normalized ratio can be considered as an indication of how close the distribution is to a normal distribution, since theoretically $f(z) = \phi(z)$ if z is normal.

Since m would almost always be large in a simulation study, the statistic would be approximately normally distributed and can also be compared against a standard normal critical value. If the test statistic is not standard normal distributed, using the approximate variances, a bootstrap estimate of the variance of the estimator must be calculated to perform hypothesis testing in practical problems. Both of these aspects are important when testing hypotheses, the distribution and the variance of the statistic.

The GPD (Johnson *et al.,* 1994) distribution is

$$F(x) = 1 - (1 + (\xi/\sigma)(x-\mu))^{-1/\xi}, x \geq \mu, \sigma > 0, \xi > 0 ,$$

$\sigma$ is a scale parameter, $\xi$ the shape parameter and $\mu$ the location parameter. The shape parameter determines how heavy the tail is and moments up to order the index $\alpha$, where $\alpha = 1/\xi$, are finite.

Using the PWM procedure means that the parameters are estimated as

$$\hat{\xi} = 2 - \hat{a}_0/(\hat{a}_0 - 2\hat{a}_1) \text{ and } \hat{\sigma} = 2\hat{a}_0\hat{a}_1/(\hat{a}_0 - 2\hat{a}_1) ,$$

$$\hat{a}_0 = \bar{x}, \hat{a}_1 = \frac{1}{n}\sum_{j=1}^{n} x_{(j)}(1-p_j), p_j = (j-0.35)/n ,$$

for a sample of size n from the two-parameter GPD and $x_{(1)} \leq ... \leq x_{(n)}$ the order statistics (Zea Bermudez, Kotz (2010)).

This section will be in two parts, a simulation testing normality for various samples sizes and shape parameters and in the second part the simulation results given in the paper of Hosking and Wallis (1987) will be used to show how previously published results can be used to check normality of three estimators in small samples.

## 3 A simulation study to test normality

In the 3.1 three ways to test normality will be used for three estimators and in 3.2 it is shown how by using a t-statistic normality can be tested when previously published simulation studies are available, where the MSE and bias of an estimator was reported.

## 3.1 Checking normality of the estimators of the shape parameter of a GPD

In this section a simulation study will be conducted to check the approximate normality of the PWM, Zhang and Stephens (2009) and two-parameter ML estimators of the shape parameter $\xi$ will be investigated. Data is simulated from the three-parameter GPD, transformed to the two-parameter by subtracting the minimum sample value.

|  | PWM | | | Zhang-Stephens | | | ML | | |
|---|---|---|---|---|---|---|---|---|---|
|  | Jarque – Bera p-value | Lilliefors p-value | p-value and t-statistic | Jarque – Bera p-value | Lilliefors p-value | p-value and t-statistic | Jarque – Bera p-value | Lilliefors p-value | p-value and t-statistic |
| | $\xi = 0.25, \sigma = 1.0$ | | | | | | | | |
| 25 | 0.0000 | 0.0000 | 0.0000 (11.8498) | 0.6380 | 0.5460 | 0.0254 (2.2390) | 0.0040 | 0.0120 | 0.0000 (11.8354) |
| 50 | 0.0000 | 0.0030 | 0.0000 (8.8430) | 0.0390 | 0.4740 | 0.0769 (1.7708) | 0.0360 | 0.3420 | 0.0000 (8.5450) |
| 100 | 0.0030 | 0.0250 | 0.0000 (6.8775) | 0.7170 | 0.7830 | 0.2178 (1.2331) | 0.8670 | 0.7640 | 0.0000 (6.0113) |
| 250 | 0.9500 | 0.8740 | 0.0000 (4.5021) | 0.4500 | 0.7030 | 0.2748 (1.0927) | 0.5340 | 0.6440 | 0.0002 (3.8053) |
| 500 | 0.0440 | 0.4960 | 0.0004 (3.5756) | 0.2760 | 0.6790 | 0.9643 (0.0448) | 0.2870 | 0.7660 | 0.0010 (3.3024) |
| | $\xi = 0.5, \sigma = 1.0$ | | | | | | | | |
| 25 | 0.0000 | 0.0000 | 0.0000 (19.7189) | 0.0000 | 0.6630 | 0.6727 (0.4225) | 0.0000 | 0.2050 | 0.0000 (9.1526) |
| 50 | 0.2050 | 0.4890 | 0.0000 (16.4923) | 0.1610 | 0.4760 | 0.7969 (0.2575) | 0.2540 | 0.7360 | 0.0000 (6.9616) |
| 100 | 0.0780 | 0.0720 | 0.0000 (12.1169) | 0.3980 | 0.8710 | 0.6917 (0.3966) | 0.4800 | 0.8700 | 0.0000 (4.4957) |
| 250 | 0.0000 | 0.0000 | 0.0000 (9.0489) | 0.3200 | 0.4950 | 0.9874 (0.0158) | 0.3740 | 0.4410 | 0.0061 (2.7493) |
| 500 | 0.0000 | 0.0040 | 0.0000 (6.6057) | 0.6250 | 0.1130 | 0.2613 (1.1240) | 0.6030 | 0.0680 | 0.4941 (0.6840) |
| | $\xi = 0.75, \sigma = 1.0$ | | | | | | | | |
| 25 | 0.0000 | 0.0010 | 0.0000 (32.8690) | 0.0000 | 0.0440 | 0.0234 (2.2705) | 0.0000 | 0.0090 | 0.0000 (7.1756) |
| 50 | 0.0250 | 0.0350 | 0.0000 (31.3081) | 0.0870 | 0.8280 | 0.0188 (2.3536) | 0.1110 | 0.9710 | 0.0000 (5.6499) |
| 100 | 0.6460 | 0.3520 | 0.0000 (30.8220) | 0.5370 | 0.4130 | 0.0020 (3.1022) | 0.5420 | 0.5710 | 0.0000 (5.0244) |
| 250 | 0.0150 | 0.0070 | 0.0000 (22.5493) | 0.0150 | 0.4040 | 0.8437 (0.1972) | 0.0150 | 0.3590 | 0.3036 (1.0293) |
| 500 | 0.0000 | 0.0000 | 0.0000 (22.5263) | 0.8010 | 0.7100 | 0.8341 (0.2095) | 0.7880 | 0.7230 | 0.5506 (0.5971) |

Table 1. testing normality of three estimators of the shape parameter of a GPD, for various sample sizes and parameters. P-values given and normality will be rejected for P< 0.05. Based on m=1000 samples generated for each sample size and set of parameters.

From the above results it can be seen that the estimator of Zhang and Stephens (2009) is approximately normally distributed especially for n>250, and the PWM not close to normal for n=500.

In figure 1 a histogram of estimated parameters using the Zhang and Stephens (2009) method is shown. The skewness and kurtosis of the three estimators PWM, Zhang&Stephens and ML are respectively: 0.2238, 3.4435, 0.0167, 3.1591, 0.0124, 3.4413.

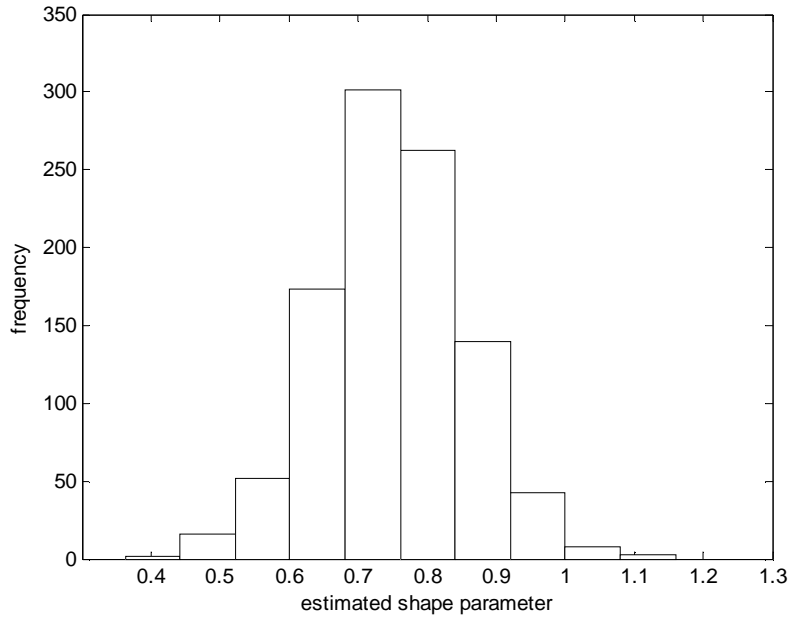

Figure 1. Histogram of m=1000, n=250 estimated shape parameter from GPD with $\xi = 0.75, \sigma = 1.0, \mu = 1.0$.

In the next simulation study the testing of hypotheses using a bootstrap estimate for the variance will be investigated. There are m=1000 simulated statistics and each time the statistic $z = (\hat{\xi} - \xi_0)/\hat{\sigma}_*(\hat{\xi})$, $\hat{\sigma}_*^2(\hat{\xi})$ the bootstrap variance for each sample. The number of rejections at the 5% level is shown.

| Testing: $H_0 : \xi = \xi_0$ | | |
|---|---|---|
| n | Average Mean, variance z | % rejected |
| $\xi_0 = 0.2, \sigma = 1.0$ | | |
| 15 | 0.0475 (0.9196) | 5.1 |
| 50 | 0.0411 (1.1021) | 6.4 |
| 100 | 0.0189 (1.2174) | 5.7 |
| 150 | 0.0229 (1.1381) | 6.8 |
| 250 | 0.0204 (1.1145) | 5.9 |

| $\xi_0 = 0.4, \sigma = 1.0$ | | |
|---|---|---|
| 15 | 0.0187 (0.9134) | 4.9 |
| 50 | 0.0859 (1.1436) | 6.7 |
| 100 | 0.0579 (1.1306) | 5.7 |
| 150 | 0.0883 (1.1467) | 7.7 |
| 250 | 0.0820 (1.1364) | 5.8 |
| $\xi_0 = 0.6, \sigma = 1.0$ | | |
| 15 | 0.2166 (1.1077) | 7.5 |
| 50 | 0.1601 (1.2994) | 8.1 |
| 100 | 0.1974 (1.2195) | 7.8 |
| 150 | 0.1195 (1.1481) | 6.9 |
| 250 | 0.1698 (1.1313) | 7.1 |

Table 2. Checking the rejection rate under the true hypothesis at the 5% level using the method of Zhang and a bootstrap estimate of the variance.

## 3.2 Using previously published results to check normality

Hosking and Wallis (1987) conducted a simulation study, generating m=50000 samples to estimate the MSE and bias of the maximum likelihood estimator (MLE), MOM and PWM estimators, tables 2 and 3 in their paper. They reported the estimated bias and RMSE, where MSE is the square of RMSE. Using the equation
$z \approx \sqrt{v} B(\hat{\theta}) / \sqrt{(MSE - B^2(\hat{\theta}))} = \sqrt{v} B(\hat{\theta}) / \sqrt{(RMSE^2 - B^2(\hat{\theta}))}$ their results will be used to check for normality in small samples. Two simulation studies where the estimator is possibly biased with different numbers of generated samples can be made comparable. Hosking and Wallis (1987) used m=50000 samples and if one wish to see if the z are of the same order size as that of a simulation study with m=1000, the statistic
$z^* \approx \sqrt{999} B(\hat{\theta}) / \sqrt{(MSE - B^2(\hat{\theta}))}$ can be calculated. Both are reported. As mentioned in section 2, the test statistic should not increase as the number of simulated samples increase if the estimator is unbiased.

| | $\sigma = 1, \xi = 0.4$ | | | | | | | | |
|---|---|---|---|---|---|---|---|---|---|
| N | ML | | | MOM | | | PWM | | |
| | Bias | RMSE | $z, z^*$ | Bias | RMSE | $z, z^*$ | Bias | RMSE | $z, z^*$ |
| 15 | 0.16 | 0.46 | 82.9561 11.7318 | 0.3 | 0.38 | 287.6118 40.6745 | 0.18 | 0.36 | 129.0995 18.2574 |
| 50 | 0.05 | 0.22 | 52.1854 7.3801 | 0.17 | 0.21 | 308.3274 43.0742 | 0.07 | 0.19 | 88.6147 12.5320 |
| 100 | 0.02 | 0.15 | 30.0828 4.2544 | 0.13 | 0.13 | 311.6512 44.0741 | 0.04 | 0.14 | 66.6667 9.4281 |

Table 3. Test statistics for normality calculated using reported bias and MSE

The estimators are not normally distributed for samples sizes less than n=100. The adjusted $z^*$ can be compared to those found in the simulation study in section 4.2 where sample sizes of m=1000 were generated and it will be seen that the test statistics calculated for the PWM estimators from the study of Hosking and Wallace (1987) are approximately equal.

## 4 A confidence interval for the log returns of the S&P500

Using the closing values of the S&P500 over the last 5 years, 2006 to 2011, 1258 log returns was calculated. A GPD was fitted to the largest 150 returns using a threshold of 0.0060. The purpose is to calculate a 95% confidence interval for $\xi$ using the above approximate normality of the Zhang and Stephens estimator and a bootstrap estimate for the variance of the estimator. The estimated values of $\sigma$ using MOM, PWM and Zhang and Stephens are 0.0046, 0.0044, 0.0044 respectively and the estimated values of $\xi$ using the three methods are 0.1494, 0.1752, 0.1919.

The bootstrap estimate of the standard deviation of the Zhang and Stephens estimate of $\xi$ is 0.0897, resulting in a 95% confidence interval of [0.0162: 0.3677], with the implication that the index of the tail observations is not less than 2.7196 (=1/0.3677). The P-P plot of the estimated distribution versus the empirical distribution function is shown in figure 4.

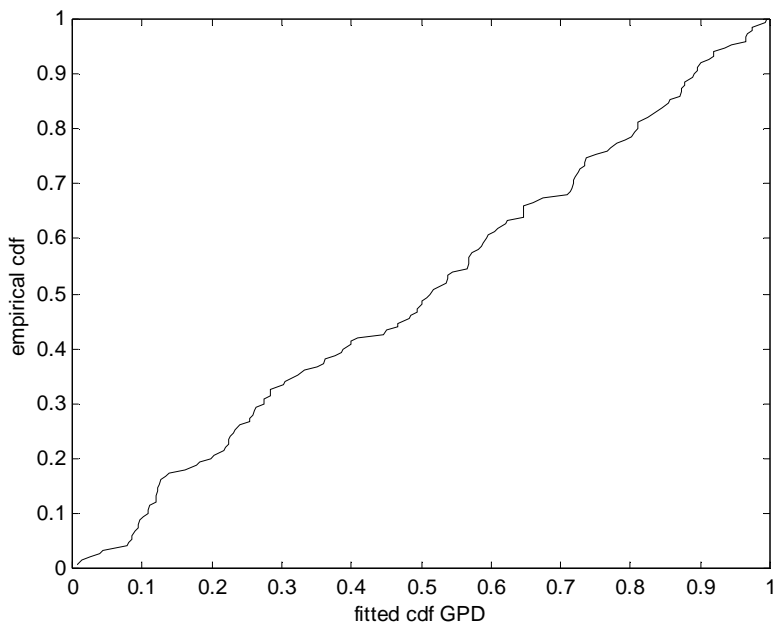

Figure 2. P-P plot of the fitted GPD and empirical distribution to the 150 largest log returns from a sample of size 1258.

It can be seen that the line is close to straight and the GPD gives a good fit.

## 5 Conclusions and Remarks

Normality in smaller samples of the estimator of the shape parameter of a GPD was investigated, and the estimator of Zhang and Stephens (2009) was shown to be close to normally distributed even in small samples, and when using excesses over a threshold when performing the estimation.

Some test are robust, especially the t-test, but the knowledge of normality, can assure more confidence in the conclusions made.

It was shown the previously published results where the estimated MSE and bias were given can be used to check the normality of estimators.

## References

De Zea Bermudez, P. , S. Kotz,  2010. Parameter estimation of the generalized Pareto distribution—Part I. *Journal of Statistical Planning and Inference*, Volume 140, Issue 6, Pages 1353-1373.

Finner, H. and Dickhaus, T., 2010. Edgeworth expansions and rates of convergence for normalized sums: Chung's 1946 method revisited. Statistics and Probability Letters, 80, 1875 – 1880.

Hosking, J.R.M. and J.R. Wallace. 1987.  Parameter and quantile estimation for the generalized Pareto distributio*n*. *Technometrics*, 29(3) , 339 – 349.

Hossain, A. , Zimmer, W. ,2003. Comparison of estimation methods for Weibull parameters: Complete and censored samples. *Journal of Statistical Computation and Simulation*, 73:2, 145 – 153.

Jarque, C.M. and Bera, A. K. (1980). "Efficient tests for normality, homoscedasticity and serial independence of regression residuals". *Economics Letters* **6** (3): 255–259.

Jarque, C.M. and Bera, A. K.  (1987). "A test for normality of observations and regression residuals". *International Statistical Review* **55** (2): 163–172.

Johnson, N.L., S. Kotz and N. Balakrishnan. 1994.  Continuous Univariate Distributions. Volume 1. Wiley, New York.

Kendall, M., Stuart, A. and Ord, J.K. (1987). *Kendall's Advanced Theory of Statistics*. Charles Griffin and Company, London.

Lilliefors, H. (June 1967), "On the Kolmogorov–Smirnov test for normality with mean and variance unknown", JASA, Vol. 62. pp. 399–402.

Zhang, J. and Stephens, M.A. (2009). A New and Efficient Estimation method for the Generalized Pareto Distribution. *Technometrics*, 51,3, 316 -325.